\documentclass[submission,copyright,creativecommons]{eptcs}
\providecommand{\event}{TFPIE 2017}
\usepackage{underscore}
\usepackage{color}
\usepackage{csquotes}

\title{Teaching Erlang through the Internet:\\ An Experience Report}
\author{Stephen Adams
\email{sa597[AT]kent.ac.uk}
\institute{School of Computing \\ University of Kent}}

\def\authorrunning{Adams}
\def\titlerunning{Teaching Erlang through the Internet}

\begin{document}

\maketitle

\begin{abstract}
Today functional programming languages are seen as a practical solution to the difficult problems of concurrent and distributed programming.   Erlang is a functional language designed to build massively scalable and fault tolerant applications. This paper describes the authors' experiences delivering a massively online open course (MOOC) on the FutureLearn platform.  
\end{abstract}

\section{Introduction}

The unexpected lifespan of Moore's Law seems to be coming to the end, soon programmers will no longer be able to rely on increases in individual processor speed to power increasingly complex software~\cite{mitMoores}. Fortunately at the same time multicore and multithreaded technology is on the rise. Imperative languages with their implicit global state and low level concurrency primitives make writing concurrent programs very difficult.

Functional programming languages offer solutions to these problems. However computer science education is predominately based around imperative languages. In 2014, of the 39 top undergraduate Computer Science programs only five taught a functional programming language to their first year students~\cite{popLanguages} and more specifically in the UK under 10\% of surveyed universities teach functional programming as the primary paradigm in their introductory courses~\cite{ukLangPop}. Many learners are turning to alternative education platforms to discover functional programming.

\subsection{Erlang}
Erlang\footnote{Named after the Danish mathematician Agner Krarup Erlang the inventor of queueing theory.} is a functional programming language designed to be massively scalable and highly fault tolerant~\cite{erlang}. It was originally developed in 1986 by Joe Armstrong, Robert Virding, and Mike Williams at the Computer Science Laboratory at Ericsson Telecom AB~\cite{erlangHistory}. Erlang's core design tenets include lightweight processes, that communicate through message passing. Erlang also boasts a ``let it fail" error handling strategy, where processes either succeed or fail and other specialised processes handle the error~\cite{armstrongThesis}.

Packaged along with the Erlang language is the Open Telecom Package or OTP. OTP is a set of libraries and design principles for developing distributed, concurrent, and fault tolerant systems~\cite{erlang}. OTP has been described as a large part of the reason Erlang is so good at these domains~\cite{learnYouErl}. 

\subsection{Erlang at the University of Kent}
Though the University of Kent's introductory programming modules are taught using Java, ``Functional and Concurrent Programming" is a mandatory stage 2 Spring term module taught since 2014. CO545's principle language is Erlang, however Haskell is also covered to a reasonable degree towards the end of the course. Erlang was chosen as the principle language because of its relatively small syntax and its message passing concurrency makes a stark contrast to Java's threads.

Additionally since teaching Erlang to undergraduates is a fairly unique offering, Kent's students are not only well trained in principles but are also capable of working with Erlang during their year in industry placements. Some companies have begun taking our students for this reason\footnote{The web security company Alert Logic for example: \url{https://www.alertlogic.com/}}.

In early 2017 the authors had the opportunity to supervise two massively open online courses (MOOC) on the online learning platform FutureLearn, Functional Programming in Erlang~\footnote{https://www.futurelearn.com/courses/functional-programming-erlang/1} which took place from February 20 to March 10 2017 and Concurrent Programming in Erlang\footnote{\url{https://www.futurelearn.com/courses/concurrent-programming-erlang/1}} (April 3 to April 21 2017). This paper will describe our curriculum and approach to teaching functional programming in a MOOC context in section~\ref{curriculum}. It will also cover our experiences teaching the courses and who participated.     

\section{Curriculum and Pedagogy}\label{curriculum}

Our first attempt at teaching a MOOC happened in the Spring of 2015 as part of the University of Kent's ``Beacon Projects"\footnote{\url{https://www.kent.ac.uk/beacon/about.html}} which were a part of the University's 50th anniversary celebrations. We ran a three weeks long pilot program, hosted through the University's Moodle site. The pilot course's materials were adapted from the first part of CO545. The pilot course involved short, two to twenty minute video lectures, along with assignments for students to complete. We also brought in Joe Armstrong co-creator of Erlang, and Francesco Cesarini the technical director of Erlang Solutions to produce ``master classes" which focused on describing how Erlang was used in the real world. The master classes were made available to the students at the end of the pilot course\footnote{See: \url{https://goo.gl/WxDHiB}, \url{https://goo.gl/RITxtd}, and \url{https://goo.gl/Wazkuc}}.

We recruited participants in the pilot MOOC from the University's current computing students and through social media associated with the Erlang community. Overall over five hundred people signed up for this pilot course. At the end of the pilot course a few things became clear based on feedback we had from our learners. Moodle, though not designed for the delivery of MOOCs did an adequate job. Most of the people we surveyed were interested in taking another MOOC course on Erlang regardless of what platform it was delivered on. At the same time learners indicated that social learning via discussion was important and a dedicated MOOC platform would have more features to support this type of learning.

Based on our experience giving this pilot MOOC we decided to develop a further three weeks of material and move our courses onto the FutureLearn platform.

Functional Programming in Erlang, the first course on FutureLearn, covered the same materials as the pilot course but we were able to expand it to use FutureLearn's support for quizzes, tests, and peer assessments. We also developed a second three week group of material, meant to follow the first course, about Erlang's concurrency and fault tolerance features.

\subsection{The curricula}

Functional Programming in Erlang was designed for someone who was familiar with at least one other programming language though not necessarily a functional one. Our goal then for this first course was to give someone unfamiliar with functional programming a basic introduction to Erlang and make them ready to tackle the second course on Erlang's concurrency features. 

The three weeks of the course are:

\begin{itemize}
	\item Getting started with Erlang
	\item Lists in Erlang
	\item Advanced functional programming
\end{itemize}

The first week began by introducing Erlang in fairly abstract ways. We covered some of its history and the ways that Erlang (as well as many functional languages) differ from imperative programming, evaluating expressions and immutability, versus procedural steps mutating a global state. We then covered the basics of Erlang's syntax, including its basic data types and how pattern matching occurred. The final activity in the first week introduced the idea of functions and recursion (both head and tail).

The second week was focused entirely on lists. Learners were first introduced to list syntax and how lists were constructed, pattern matched, and used in functions. Once they became familiar with the list syntax we looked at common strategies for defining Erlang functions. Finally they were tested on their knowledge with an assessment.

The final week introduced the learners to higher-order functions and the week's second activity involved modelling the game rock-paper-scissors. The final activity of the course linked to the master classes so that people could continue their learning if they wanted and asked what people's general opinion of Erlang was.

Following a three week break the second course on concurrency was run. Like the first course Concurrent Programming in Erlang was three weeks long. This course's outline is:

\begin{itemize}
	\item Concurrency - nuts and bolts
	\item Concurrency - making code robust		
	\item Scaling it up
\end{itemize}

The first two weeks cover Erlang's built in features and how these features affect the design of concurrent systems. The third week on scaling systems introduces the learners to the Open Telecom Platform (OTP) the set of middleware libraries for developing highly distributed systems that is included with Erlang~\cite{erlang}.

The first week explained Erlang's concurrency system, the actor model. Processes are lightweight to create and destroy and communicate with each other through message passing. Erlang's messaging system is asynchronous, each process has a mailbox and will handle messages in the order they are received. The second week then covered Erlang's unique error handling philosophy, ``let it fail." Erlang processes are expected to either succeed and if they can't they should fail immediately without doing any error handling~\cite{armstrongThesis}. Specialized supervisor processes should be designated to monitor other processes. When children processes fail the supervisors can then take the appropriate action, such as error reporting or simply restarting the failed process.

The course finished with a bit of material about OTP. Many of the concurrent design patterns we had our students implement in the first two weeks of the course are actually standard templates that OTP defines.

\subsection{FutureLearn's Course Creation}

FutureLearn provides many ways to configure and deliver online courses. Every FutureLearn course is broken down by week, and each week is further broken down into ``activities." Each week is used to suggest the pace that learners should aim for, and activities organise steps. Steps are the smallest section of a course\cite{flPartners}.

There are several different step types that a course is built from. Each step typically hosts some materials and a discussion section for learners to have a conversation about that step's content. Broadly speaking steps can be divided into two different types which could be described as teaching steps and doing steps. A teaching step's primary purpose is to convey information to the learners, textual articles and audio/video steps are examples of this. Doing steps are more interactive. They actively work to engage learners to do something, whether that be participate in a discussion, take a multiple choice quiz or test\footnote{Quizzes allow for infinite attempts whereas learners only have three attempts at each test question}, do an exercise, or submit an assignment for peer review and review other learner's work. Every step type is designed to support active learning. Doing steps requires some learner interaction to pass, however, in teaching steps the learner may optionally engage in the discussion but this isn't required to complete the step.

\section{Who participated}

In the end 5,642 people enrolled in Functional Programming in Erlang and 1,965 people enrolled in Concurrent Programming in Erlang. The majority of the learners came from the UK and the USA but in total 149 countries were represented.

\begin{table}
\begin{center}
\begin{tabular}{| l | l | l |}
\hline
\textbf{Continent} & \textbf{FP Erlang} & \textbf{Concurrent Erlang} \\ \hline
Africa & 8 & 4 \\ \hline
Asia & 20 & 15 \\ \hline
Australia & 2 & 2 \\ \hline
Europe & 45 & 50 \\ \hline
North America & 18 & 21 \\ \hline
South America & 7 & 8 \\ \hline
\end{tabular}
\caption{Percentage of enrolment by continent}
\label{continents}
\end{center}
\end{table}

Where users were located was determined based on their IP addresses. A survey was also sent out to everyone who enrolled on the course and from the people who responded we can roughly determine the ages of our learners as seen in table~\ref{ages}.

\begin{table}
\begin{center}
\begin{tabular}{| l | l | l |}
\hline
\textbf{Age Range} & \textbf{FP Erlang} & \textbf{Concurrent Erlang} \\ \hline
\textless 18 & 1 & 0 \\ \hline
18-25 & 18 & 12 \\ \hline
26-35 & 28 & 27 \\ \hline
36-45 & 19 & 22 \\ \hline
46-55 & 16 & 20 \\ \hline
56-65 & 10 & 12 \\ \hline
\textgreater 65 & 5 & 4 \\ \hline
\end{tabular}
\caption{Percentage of enrolment by age}
\label{ages}
\end{center}
\end{table}

\subsection{Prior experience}

Beyond the basic demographic information we prompted the learners to introduce themselves on the first step of each course and to briefly describe why they were taking the course what their previous programming experience was. From all the responses we discovered that our learners were familiar with over 40 different programming languages. The top ten languages that people claimed to have used before are shown in table~\ref{progLangs}\footnote{The percentages in table~\ref{progLangs} sum to greater than 100\% due to single respondents having experience with multiple programming languages.}.  

\begin{table}[t]
\parbox{.45\linewidth}{
\centering
\begin{tabular}{| l | l |}
\hline
\textbf{Language} & \textbf{\% of Respondents} \\ \hline
C/C++ & 28  \\ \hline
Java & 28  \\ \hline
JavaScript & 26  \\ \hline
Python & 22  \\ \hline
Ruby & 17 \\ \hline
Haskell & 14  \\ \hline
PHP & 12  \\ \hline
C\# & 8  \\ \hline
Erlang & 6  \\ \hline
Scala & 6  \\ \hline
\end{tabular}
\caption{Ten most popular languages}
\label{progLangs}
}
\hfill
\parbox{.45\linewidth}{
\centering
\begin{tabular}{| l | l |}
\hline
\textbf{Language} & \textbf{\% of Respondents} \\ \hline
Haskell & 14  \\ \hline
Erlang & 6  \\ \hline
Scala & 6  \\ \hline
Elm & 5  \\ \hline
Clojure & 4  \\ \hline
F\# & 2  \\ \hline
Scheme/Racket & 2  \\ \hline
ML/OCaML & 1  \\ \hline
Idris & \textless 1  \\ \hline
Emacs Lisp & \textless 1  \\ \hline
\end{tabular}
\caption{Functional Language Popularity}
\label{funcLangs}
}
\end{table}

Our learners' experience seems very much in line with broader industry trends. Seven of the languages in table~\ref{progLangs} also appeared in the ten most popular programming languages in the May 2017 TIOBE index~\cite{tiobe}. The three exceptions to this are Haskell, Erlang, and Scala who's TIOBE ranks are 38th, 41st, and 31st respectively. 

It makes sense that our learners would have a significant interest in functional programming languages prior to joining our course. Table~\ref{funcLangs} shows the number of people with experience in functional languages only. A very common reason why learners wanted to take our course was that they wanted to improve their understanding of functional programming techniques independently of the language the course was taught in.  

\section{What happened}

This section describes our experience running both of these courses. We had previously ran the pilot MOOC so we weren't completely new to the concept of MOOCs but this was our first time using the FutureLearn platform in particular and these courses were an order of magnitude larger than our pilot course.

\subsection{Participation}

If you don't have experience running or taking an online course before it may seem unmanageable for two people to run a course with over 5,000 people in it, but of those people enrolled only a fraction even begin the course and a fraction of those finish it. FutureLearn categorises learners by how far they progress through and how they interact with the course. The categories are: 

\begin{itemize}
	\item Joiner - Total number of people enrolled on the course
	\item Learner - Joiners who have at least viewed one step
	\item Active Learner - Learners who have completed at least one step
	\item Social Learner - Learners who have left at least one comment
	\item Fully Participating Learners - Learners who have completed all tests and at least 50\% of the steps
\end{itemize}

We have summarized how many people fit into each of these categories for our two courses in table~\ref{learnCats}.

\begin{table}[h]
\begin{center}
\begin{tabular}{| l | l | l | l |}
\hline
\textbf{Learner Category } & \textbf{FP in Erlang} & \textbf{Concurrent Programming} & Notes \\ \hline
Joiners & 5,642 & 1,965  \\ \hline
Learners & 3,858 & 1,117 & 68\% \& 57\% of joiners \\ \hline
Active Learners & 2,683 & 676 & 70\% \& 61\% of learners \\ \hline
Social Learners & 586 & 142 & 15\% \& 13\% of learners \\ \hline
Fully Participating Learners & 374 & 40 & 10\% \& 4\% of learners \\ \hline
\end{tabular}
\caption{Learners in each category}
\label{learnCats}
\end{center}
\end{table}

Even though relatively few people comment it's still too much for our small team to respond to every question. We actively encourage learners to answer each other's questions and give informal feedback about each other's exercises. We would only comment when something needed to be said about the course materials or there seemed to be a widespread misunderstanding though this was fairly rare. 

One thing we noticed from the pilot MOOC is that it was difficult for us to participate in discussions without stopping the discussion\footnote{Maybe people felt that our comments ``settled" the issue?}. This time we decided that at the end of each week we would record a short video talking about issues that came up throughout the previous week. This meant that the discussion sections were for learners to interact with each other and our thoughts would come through a different medium. 

\subsection{Learner Feedback}

After each exercise and assessment and in the final step of both of the courses we requested feedback so that we can continue to improve the materials for these courses and our approach to MOOC creation in general. People were more than happy to give us feedback at most opportunities.

\subsubsection{Workload}

Both of the course descriptions estimated that learners should spend about five hours a week to complete the course. It seems that most people spent a lot more time than that. Interestingly this was seen as a positive thing during the functional programming course but more of a negative thing during the concurrency course.

When designing the second course we wanted to make the exercises much more open ended so that learners could explore more. Many of the assignments and examples in the second course were centered around a ``frequency server" example. The server would allocate and deallocate numbers when they were requested through message passing. Throughout the course the frequency server was expanded through both the video lectures and student's exercises. 

During the second week we wanted the learners to attempt to build their own version of a supervisor process that would clean up the system if the server failed. Up to this point we had spoken in fairly abstract terms about supervisors. 

The feedback we got was that this exercise was much too open ended. Students were unsure what we really wanted and without previous concrete examples of supervisors they were left aimlessly trying things not sure what we wanted. One learner described their experience as:

\begin{displayquote}
``I wasted a colossal amount of time trying to come up with a 'hardened frequency server'. I wish there had been a warning, something like: "Don't spend more than 5 hours on this exercise". I didn't realise that there wasn't going to be a good answer...'
\end{displayquote} 

This seemed to be the key difference between the workload feedback we received in both courses. The first course had a high workload but learners felt that they were always working on the right thing, it just took longer than expected (they just asked that we update our estimates). For the second course learners didn't enjoy just trying things without knowing if what they had was a good solution which caused a great deal of frustration.

\section{Conclusion}

Overall people were very positive about their experiences with the course. This is of course tainted by survivorship bias because we hear from people who completed the course (or at the very least looked at the final step). The percentage of learners who ended up in each category from our courses was lower than the average all FutureLearn courses as seen in table~\ref{compPercents}.

\begin{table}[h]
\begin{center}
\begin{tabular}{| l | l | l | l |}
\hline
\textbf{Category} & \textbf{FutureLearn Average} & \textbf{FP Average} & \textbf{Concurrency Average} \\ \hline
Learners & 50\% of Joiners & 68\% & 57\% \\ \hline
Active Learners & 81\% of Learners & 70\% & 61\% \\ \hline
Social Learners & 38\% of Learners & 15\% & 13\% \\ \hline
Fully Participating Learners & 21\% of Learners & 10\% & 4\% \\ \hline
\end{tabular}
\caption{Average number of learners in each category}
\label{compPercents}
\end{center}
\end{table}

Our courses had more people become learners (those that at least looked at a step) than FutureLearn as a whole but the number of people who went on to either complete a step (active learners) or comment at least once (social learners) was lower than FutureLearn's averages. This may mean that the workload is turning people away.

\subsection{Recommendations for other MOOC organisers}

For other educators who are interested in teaching their own online courses I would offer a few recommendations in this section. The first recommendation is that setting the learners expectations is key. We were very upfront with learners that because there were only two educators working on the course we could not be relied upon to answer most questions. Instead we asked our students to help each other and we would chime in when we could. Being honest and up front with our students allowed us to 

\subsection{Future Work}

The functional programming course was popular enough that many people requested we run it again. We have scheduled another three week offering of the course to begin on the 29th of May 2017. 

A much requested addition to our course was more information on the OTP framework. Given that both our courses had high workloads we have begun planning a third course specifically about OTP. We are currently investigating developing this course along with an industrial partner and we are discussing with FutureLearn about incorporating these three Erlang courses into a single ``program." 

Today functional programming is being seen, more and more, as a practical solution to design of highly distributed and concurrent programs. Many traditional educational institutions are still teaching imperative programming first and so people are turning to alternative opportunities to augment their knowledge. In this paper we have described our methodology and experience teaching two massively online open courses with the functional programming language Erlang.

\section*{Acknowledgements}
The authors would like to thank Mark O'Connor, Distance Learning Technologist at the University of Kent for his partnership throughout this entire enterprise. His advice and mentoring on preparation and recording as well as editorial and production help made these courses possible. We also would like to thank Claire Lipscomb our contact at FutureLearn for her dedication and prompt responses to our questions.

Finally we must thank all of the people who have participated in all of our MOOCs. Their hard work and honesty is much appreciated.

\bibliographystyle{eptcs}
\bibliography{tfpie}
\end{document}